\documentclass[twocolumn,showpacs,prb,amsmath,amssymb,floatfix,eqsecnum]{revtex4}

\usepackage{dcolumn}
\usepackage{amsmath,amssymb}
\usepackage{bm}
\usepackage{graphicx}
\usepackage{color}
\newcommand{\bd}{\bm}

\newcommand{\eins}{\mbox{$1 \hspace{-1.0mm} {\bf l}$}}

\begin{document}

\title{Interplay of topology and geometry in frustrated 2d Heisenberg
magnets}
\author{N.~Hasselmann$^1$}
\author{A.~Sinner$^2$}
\date{August 22, 2004}

\affiliation{$^1$Max-Planck-Institute for Solid State Research, Heisenbergstr.~1, D-70569 Stuttgart, Germany\\
$^2$Institut f\"ur Physik, Universit\"at Augsburg, D-86135 Augsburg, Germany}

\begin{abstract}
We investigate two-dimensional frustrated Heisenberg magnets using
non-perturbative renormalization group techniques. These magnets allow
for point-like topological defects which are believed to unbind and drive
either a crossover or a phase transition  
which separates a low temperature, spin-wave dominated
regime from a high temperature regime where defects are abundant.
Our approach can account for the crossover qualitatively and
both the temperature dependence of the
correlation length as well as a broad but well defined peak
in the specific heat are reproduced. We find no signatures
of a finite temperature transition and an accompanying diverging
length scale. Our analysis is consistent
with a rapid crossover driven by topological defects.

\end{abstract}

\maketitle

\section{Introduction}
Frustrated magnets have a number of highly fascinating properties which
have been the focus of intense research interest for some time. 
These include  magnets which do not order but where a 
macroscopic
number of competing low lying states give rise to strong correlations and a large
low-temperature entropy, or spin-liquids where quantum
fluctuations prevent ordering and exotic quasi-particles appear, 
see Ref.~[\onlinecite{Balents10}] for a recent review.
A much simpler situation arises in
classical magnets if the frustration
is not sufficiently strong to prevent an ordered ground state. In this case
the ground state has a broken symmetry and the low temperature  
excitations are just
spin waves.
 However, even 
classical frustrated magnets which do order are not
completely understood, which can be attributed to a large part
to a non-trivial order parameter which characterizes such magnets. 
In $d=2$, as was first pointed out by Kawamura and Miyashita,\cite{Kawamura84} 
the order parameter
manifold of a frustrated Heisenberg magnet allows for point-like
topological $Z_2$ defects and the influence of these defects
on the properties of the magnet at finite temperatures
proved very difficult to quantify. 
In $2d$ collinear XY magnets topological defects are responsible for the
Berezinskii-Kosterlitz-Thouless (BKT) transition. However, in this
case the perturbative $\beta$-function of the XY coupling constant,
which is sensitive only to the geometry but not the topology of the
order parameter space, 
vanishes. This is very different from the situation in frustrated Heisenberg
magnets.

The major difficulty with $2d$ frustrated Heisenberg magnets is 
the combined presence of both point-like defects, originating from the
topological properties of the order parameter space,\cite{Kawamura84} 
and the phenomenon of
asymptotic freedom which has its root in the local geometry of
the order parameter space. In contrast to collinear XY magnets, where
the Villain approximation allows to map the problem on the $2d$ Coulomb
gas which can be well studied using RG techniques,\cite{Jose77} no similar
tool is available for frustrated Heisenberg models.
The yet unsolved
question is whether or not a finite temperature transition exists in
frustrated Heisenberg magnets. In particular, for the simplest such
model,  the Heisenberg antiferromagnet on a triangular lattice (HAFT),
this question has been addressed repeatedly over the years, without
a definite conclusion.
Monte Carlo (MC)
simulations of the HAFT
have found indications of a vortex unbinding at a finite 
temperature $T_{\rm cross}$.\cite{Southern95,Wintel95,Kawamura10} 
The vorticity modulus, which measures
the response of the magnet to an imposed twist along a path
which encloses a vortex core, has been shown to vanish\cite{Southern95}
 at $T_{\rm cross}$. 
Further indications of a finite temperature transition
can be found from the phase diagram of the HAFT in a magnetic 
field.\cite{Shannon11,Mike11} In small fields, there are two finite temperature
transitions. There is a BKT 
transition from a low temperature canted state with quasi-long-range
order of the transverse spin-components
to an intermediate state which has a vanishing
spin stiffness. A second transition at higher temperatures restores 
the sub-lattice symmetry of the magnet, which is broken in both
low temperature phases. It is unclear from MC what happens in the
zero field limit, but both transitions are of the 
order of $T_c\approx 0.3 J$ (where $J$ is antiferromagnetic
exchange constant)
for very small fields, a similar temperature to where at zero field a
vortex unbinding seems to occur. In a perturbative RG analysis
some indication of a fixed point in $d=2$ which might correspond 
to a topological phase transition were reported,\cite{Calabrese01} 
see however also Ref.~[\onlinecite{Delamotte10}]. Experimentally,
there are also several reports on indications of a vortex driven 
transition.\cite{Olariu06,Yaouanc08,Yamaguchi08,Schmidt13}

Perhaps the cleanest demonstration
of the role of topology comes from a comparison of MC simulations of
two different matrix models representing
interacting tops, which both share the same geometrical
properties but
differ in their topology.\cite{Caffarel01}
The model which allows for topological defects
shows a clear finite temperature peak in the specific heat and
a crossover in the correlation length dependence on $T$ which are
both absent in the topological trivial model.

The properties of the long wavelength modes of the magnet
is described by a non-linear $\sigma$-model (NL$\sigma$M).
The order parameter space for a frustrated Heisenberg magnet has
the symmetry 
$SO(3)\times SO(2)/SO(2)\sim SO(3)$, 
see e.g. Ref.~[\onlinecite{Delamotte04,Kawamura98}]
for a discussion of the symmetries.  
While this model 
describes well the physics of the Heisenberg AF on the
triangular lattice at low temperatures \cite{Southern93,Wintel95},
its  perturbative $\beta$ function is 
not sensitive to
the
topological properties of $SO(3)$ which has 
a nontrivial homotopy group\cite{Kawamura84} $\pi_1[SO(3)]=Z_2$ and thus allows
for topological defects which could be generated either
through temperature or disorder.\cite{Hasselmann04}

An alternative continuum model for frustrated magnets is based
on a Landau-Ginzburg action which includes also massive excitations.
The advantage of using a Landau-Ginzburg model in conjunction
with a non-perturbative RG (NPRG) approach is its ability to
describe  the BKT transition of the
$2d$ XY model,  
without
relying on a mapping to the Coulomb gas.\cite{Gersdorff01}
Although it is not
well understood  how exactly topology enters the NPRG flow, its success
in the study of the XY model makes the NPRG a promising
approach to
the physics of $Z_2$ defects in frustrated Heisenberg models.
Here, we follow this ansatz and present results for $d=2$. 

In Sec.~\ref{sec:model} we discuss the different field theoretical
approaches to the HAFT and 
present the Landau-Ginzburg model which we investigate here.
Although the Landau-Ginzburg model applies to non-collinear ordered
magnets in general, we shall concentrate
here on the HAFT model in our numerical analysis 
and estimate appropriate initial
values for the NPRG 
in Sec.~\ref{sec:model}. 
The  NPRG approach is presented in Sec.~\ref{sec:NPRG},
and the approximation of the effective average
action are presented and discussed in Sec.~\ref{subsec:local} and
\ref{subsec:nonlocal}. The derivation of the
 flow equations 
is discussed in Sec.~\ref{subsec:deriv}. 
Results for the NPRG approximation of the HAFT model are 
presented in Sec.~\ref{sec:results}, where we calculate
both the temperature dependence of the spin correlation
length and the specific heat. Our results show a clear
crossover behavior of the temperature dependence
of the correlation lenght, from a low temperature
exponential dependence characteristic as it is also
obtained within a NL$\sigma$M approach, to a much
weaker temperature dependence at higher temperatures.
This crossover is also visible as a broad but well
defined peak at the crossover temperature
in the specific heat. We stress that while this crossover has been
repeatedly observed in MC data, it is not captured
by the NL$\sigma$M and it also has not
yet been successfully described by other analytical approaches.
We close with a summary
in Sec.~\ref{sec:summary}.

\section{The antiferromagnetic Heisenberg model on the triangular lattice}
\label{sec:model}

We concentrate on 
one of the simplest frustrated Heisenberg models, the 
Heisenberg antiferromagnet on a triangular lattice (HAFT). It is defined by
\begin{align}
   {\cal H}=J\sum_{< i,j>} {\bf S}_i \cdot {\bf S}_j \, ,
\end{align}
where the sum is over nearest neighbors of the triangular lattice, ${\bf S}_i$
are three component unit vectors with ${\bf S}_i^2=1$, and $J>0$.
The zero temperature ground state is the well known planar $120^\circ$ state,
where neighboring spins have angles $\pm 120^\circ$.

MC simulations \cite{Southern93}
have convincingly demonstrated that at low temperatures
the $2d$ HAFT model is
well described by a NL$\sigma$M which has the form
\begin{align}
  S=\frac{1}{2}\int_x \sum_{i=1}^3 p_i (\partial_\mu {\bd n}_i)^2 \, ,
\label{eq:nlsm}
\end{align}
where the ${\bm n}_i$ are orthonormal three-component unit vectors and the
$p_i$'s are three stiffnesses (divided by the temperature), and
$\int_x=\int d^d x$. 
Because of the 
planar spin orientation in the ground state one has $p_1=p_2$ which holds
both at 
the bare level but also throughout the renormalization group flow.

The alternative Landau-Ginzburg approach for frustrated magnets has been developed early
on, see e.~g.~Ref.~[\onlinecite{Kawamura88}], and has usually been applied
to study frustrated magnet close to $d=4$. 
It has also been the basis
of a thorough non-perturbative RG (NPRG) analysis\cite{Tissier00,Delamotte04}
where flow equations were derived for all $2< d < 4$. The central
functional in the NPRG approach is the effective average action
which is also the generating functional of one-particle
irreducible correlation functions, and the NPRG provides a framework
in which the flow of this functional connects the bare effective
average action, which is identical to the bare action, to the
fully renormalized generating functional of irreducible 
vertices.\cite{Berges02,Kopietz10}
The simplest
approximation for the
effective average action used in the study of
frustrated magnets has the form\cite{Tissier00,Delamotte04}
\begin{align}
  \Gamma_\Lambda[{\bd \Phi}_1,{\bd \Phi}_2]&=\int_x  \Big\{ \frac{Z_\Lambda}{2}
  \big[ (\partial_\mu {\bd \Phi}_1 )^2 
  +(\partial_\mu {\bd \Phi}_2 )^2 \big]
  \nonumber \\
  &
  +\frac{\lambda^0_\Lambda}{4}\big[ \rho/2 -\kappa_\Lambda \big]^2 +
\frac{\mu^0_\Lambda}{4} \tau \nonumber
\\
& + \frac{\Omega_\Lambda}{4} \left( {\bd \Phi}_1 \cdot \partial_\mu {\bd \Phi}_2- 
{\bd \Phi}_2 \cdot \partial_\mu {\bd \Phi}_1 \right)^2
\Big\}
\label{eq:EffActDE}
\end{align}
where $\rho={\rm Tr} {^t}\Phi \Phi$ and $\tau=(1/2) {\rm Tr} [ {^t}\Phi \Phi-\eins
\rho/2]^2$ are local  invariants of the theory. Here, the symmetry
$SO(3)\times SO(2)$ for Heisenberg ($N=3$) models has
been generalized for general $N\geq 2$ to a $O(N)\times O(2)$ symmetry and
the symmetry of the symmetry broken ground state
is $O(N-2)\times O(2)$. 
The subscript $\Lambda$
in $\Gamma_\Lambda$ indicates that all parameters entering (\ref{eq:EffActDE})
depend on the cutoff scale $\Lambda$. The fields ${\bd \Phi}_{1,2}$ have
$N$ components (the same number of components as the lattice spins),
are 
orthogonal in the ground state and span
the planar order of a frustrated magnet,\cite{Delamotte04} 
such as e.g. the $120^\circ$ state of the
HAFT.
Further, $\Phi=({\bd \Phi}_1,{\bd \Phi}_2)$ is
a $2\times N$ matrix such that 
\begin{equation}
  {^t}\Phi \Phi=\left(
    \begin{array}{cc} 
      {\bd \Phi}_1 \cdot {\bd \Phi}_1 &   {\bd \Phi}_1 \cdot {\bd \Phi}_2 \\
      {\bd \Phi}_2 \cdot {\bd \Phi}_1  & {\bd \Phi}_2 \cdot {\bd \Phi}_2
    \end{array}
  \right) .
\end{equation}
Thus, one has the expressions $\rho={\bd \Phi}_1^2 +{\bd \Phi}_2^2$ 
and $\tau= \big({\bd \Phi}_1^2 -{\bd \Phi}_2^2\big)^2/4 + \left({\bd \Phi}_1
  \cdot {\bd \Phi}_2\right)^2$. Both $\lambda_\Lambda^0$ and 
$\mu_\Lambda^0$ are positive coupling
parameters, where $\lambda_\Lambda^0$ controls the magnitude of the vector fields
and $\mu_\Lambda^0$ ensures that ${\bd \Phi}_1$ and ${\bd \Phi}_2$ are orthogonal
in the ground state. If both $\lambda_\Lambda^0$ and $\mu_\Lambda^0$ 
become very large, ${\bd \Phi}_1$ and  ${\bd \Phi}_2$ are forced into
a configuration where they are orthogonal
with fixed length and, for $N=3$, can be identified
with the ${\bd n}_1$ and ${\bd n}_2$ fields of the NLSM, 
after a suitable rescaling
such that both ${\bd \Phi}_1$, ${\bd \Phi}_2$ have norm one. 
The third 
field ${\bd n}_3$ of the NL$\sigma$M 
is not independent of ${\bd n}_1$ and ${\bd n}_2$ 
but fixed
by the relation
${\bd n}_3={\bd n}_1 \times {\bd n}_2$. However, to recover correctly 
the three independent fluctuation terms $(\partial_\mu {\bd n}_i)^2$
 of the  ${\bd n}_i$ fields within the
Ginzburg-Landau model (\ref{eq:EffActDE}), it is necessary
to add the $\Omega_\Lambda$-derivative term, which is the
 only derivative term at fourth order in the fields which
directly renormalizes 
the gapless modes of the model.\cite{Azaria90,Tissier00,Delamotte04} 

A central role is played by the parameter $\kappa_\Lambda$
which is the order parameter of the theory. It gives the magnitude
of the ordered magnetization (the canted 120$^\circ$ magnetization)
around which $\rho$, which corresponds
to the local magnetization, fluctuates. It is initially finite,
since the IR modes are cut off, but the further the IR cutoff 
$\Lambda$ is
reduced, the stronger $\kappa_\Lambda$ is suppressed (for $d=2$).
The vanishing of $\kappa_\Lambda$ at some finite scale $\Lambda$
signals the absence of 120$^\circ$ order and the spin-correlation
length is then determined by $2\pi/\Lambda$. 

For the case considered here, the triangular AF, we have $N=3$ and
the fields ${\bd \Phi}_{1,2}$ can be locally
related to the microscopic spins of the triangular AF. This is
done by partitioning the spins first into plaquettes of three spins,
where each of the spins belongs to one of 
the three sublattices
associated with a 120$^\circ$ order. We then have\cite{Azaria93}
\begin{subequations} 
\begin{align}
\frac{3}{\sqrt{2}} {\bd \Phi}_1 & =-\frac{1}{2}(\sqrt{3}+1){\bd S}_1 + \frac{1}{2}(\sqrt{3}-1){\bd S}_2 +{\bd S}_3 \, , \\
\frac{3}{\sqrt{2}} {\bd \Phi}_2 & = \frac{1}{2}(\sqrt{3}-1){\bd S}_1 - \frac{1}{2}(\sqrt{3}+1){\bd S}_2 +{\bd S}_3 \, ,
\end{align}
\label{eq:spinmicro}
\end{subequations}
where ${\bd S}_1 \dots {\bd S}_3$ are the three spins of
 a local triangular plaquette.
Note that we have for three spins six degrees of freedom,
the same number as we have in the 
two unconstrained three-component fields ${\bm \Phi}_1$ and ${\bm
  \Phi}_2$. 
One can easily check that 
${\bd \Phi}_1\cdot {\bd \Phi}_2 = (2/9)(2{\bd S}_1\cdot{\bd S}_2- {\bd S}_1\cdot{\bd S}_3
-{\bd S}_2\cdot{\bd S}_3)$ and 
$({\bd S}_1+{\bd S}_2+{\bd S}_3)^2=9(1-{\bd \Phi_1}^2/2-{\bd \Phi_2}^2/2)$
which both vanish  in the perfectly ordered 120$^\circ$
ground state in which
the fields are chosen to be normalized such that
${\bd \Phi}_{1,2}^2=1$. 
As we discuss in more detail below, the model defined by Eq.~(\ref{eq:EffActDE})
supports $2N$ modes of which $2N-3$ are gapless at $T=0$.
There are two 
modes with gaps  $\kappa_\Lambda \mu_\Lambda^0$ and one with a mass $\kappa_\Lambda \lambda_\Lambda^0$.
At any finite temperature all modes eventually become gapped, however at very small temperatures
the IR physics is completely dominated by 
the $2N-3$ modes which are initially gapless. This low temperature regime is well
described by a NL$\sigma$M. In principle it would also be possible to
start our investigation from the paramagnetic phase which has $\kappa=0$,
however, it is then far more difficult to ensure that the symmetries of the
model are not violated in the flow. Thus, within the same spirit as
in the NL$\sigma$M approach, we assume a local order and investigate how this
order is destroyed by fluctuations.

The relation between the NL$\sigma$M and the NPRG approach near
$d=2$ (and for any $N$) has been
established in Ref.~[\onlinecite{Delamotte04}] who showed that in the limit
of large masses the NPRG reduces to 
\begin{subequations}
\begin{align}
  \partial_\ell \eta_1= -(d-2)\eta_1 +N-2 
-\frac{\eta_2}{2\eta_1} \, , \\
  \partial_\ell \eta_2= -(d-2)\eta_2+\frac{N-2}{2} 
\Big(\frac{\eta_2}{\eta_1}\Big)^2 \, ,
\end{align}
\label{eq:oneloop}
\end{subequations}
with $\ell=-\ln \Lambda/\Lambda_0$ and
\begin{subequations}
  \begin{align}
    \eta_1&=2 \pi \tilde{\kappa}  
\label{eq:eta1}
\\
    \eta_2&=4 \pi \tilde{\kappa} (1+\tilde{\kappa} \tilde{\Omega})
    \, ,
    \label{eq:eta2}
  \end{align}
\end{subequations}
where we introduced the rescaled dimensionless parameters
\begin{subequations}
\begin{align}
  \tilde{\kappa}= Z_\Lambda \Lambda^{2-d} \kappa_\Lambda \, , 
\label{eq:kapparescaled}
\\
  \tilde{\Omega}= Z_\Lambda^{-2} \Lambda^{d-2} \Omega_\Lambda \, .
\label{eq:omegarescaled}
\end{align} 
\end{subequations}
These reproduce for $N=3$ the one-loop $\beta$-functions of the stiffnesses 
entering the NL$\sigma$M given in Eq.~(\ref{eq:nlsm}) if one identifies $\eta_1/2=p_3+p_1$ and $\eta_2/4=p_1$.
One important prediction of these RG equations (which is preserved
also at two-loop order\cite{Azaria92}) is an interaction driven
enhancement of symmetry. This can be expressed by 
the parameter 
$\alpha=(p_1-p_3)/(p_1+p_3)$ which flows towards the fixed point $\alpha^*=0$,
i.e. all the $p_i$'s become asymptotically equal in the IR limit 
$\ell\to \infty$.
This signals an enhancement of the original symmetry to $O(4)/O(3)$
and this symmetry determines the critical behavior at finite $\epsilon$ in
a $d=2+\epsilon$ expansion. We emphasize that this 
enhanced symmetry is however only expected
at low temperatures and in the IR limit.

Dombre and Read [\onlinecite{Dombre89}] 
derived
the values of the $p_i$'s of the NL$\sigma$M (\ref{eq:nlsm}) appropriate for
the HAFT at the original lattice scale and found
$p_1=p_2\approx \sqrt{3} J/4 T$ and
$p_3\approx 0$. This derivation was based on a local rigidity constraint where
the spins were grouped into local three-spin plaquettes within which they where
assumed to be rigid. We will use these values to fix
the derivative terms $Z_{\Lambda_0}$ and $\Omega_{\Lambda_0}$ in our initial 
effective action.

While rigid rotations of the spins within a plaquette account for the
three initially gapless modes,  we can easily understand also the nature of
the three gapped modes from looking at a single
plaquette if we relax the rigidity constraint.
For a local three-spin plaquette we have
\begin{align}
  {\bf S}_1 \cdot {\bf S_2}+{\bf S}_1 \cdot {\bf S_3}+{\bf S}_2 \cdot {\bf S_3}
 &= {\bm L}^2/2-3/2 \, ,
\end{align}
where ${\bm L}={\bf S}_1 +{\bf S}_2 +{\bf S}_3$ is the ferromagnetic moment
of the three spins, which vanishes for the planar 120$^\circ$ ground state.
Small fluctuations around that state give rise to two  massive excitations with
energy $3J/4$ and a singlet with excitation $3J/2$. This is the same
structure of massive modes which we obtain from Eq.~(\ref{eq:gamma2})
which has two modes with mass $\kappa_\Lambda \mu^0_\Lambda$
and one with mass $\kappa_\Lambda \lambda^0_\Lambda$. 
Dividing by the size
of the unit cell $a^2 \sqrt{3}/2$ (where $a$ is the nearest neighbor
distance) we thus estimate 
$\mu_{\Lambda_0}^0 \kappa_{\Lambda_0}=\beta J \sqrt{3}/2$ and 
$\lambda_{\Lambda_0}^0
\kappa_{\Lambda_0} =\beta J \sqrt{3}$ in units such that $a=1$ and 
where $\Lambda_0$
is the UV cutoff of the model which originates from the lattice. We
fix it 
by matching it with the smallest wavevector in the (magnetic)
Brillouin zone boundary,\cite{Kawamura88} $\Lambda_0\approx  2\pi/3a$.
Since we normalized the ${\bd \Phi}_i$ fields to be equal
to one in the zero temperature ground state,
we set the initial normalization of the ${\bd \Phi}_i$ equal to one by
choosing $\kappa_{\Lambda_0}=1$. We finally rescale the fields
to have the initial value $Z_{\Lambda_0}=1$, the initial value
of $\Omega_{\Lambda_0}$ is zero.

While the switch to a continuum
field theory is necessarily only approximate, with this estimate
of initial values of the coupling constants 
we nonetheless
expect to get reasonable approximate values for the
relevant energy scales of the model.

\section{Nonperturbative renormalization analysis of the
Landau-Ginzburg model}
\label{sec:NPRG}
The model defined by Eq.~(\ref{eq:EffActDE}), and extensions thereof including all local terms up to 10th order
in the fields as well as two more additional derivative terms of fourth order,
were investigated  in Refs.~[\onlinecite{Tissier00,Delamotte04}]. The
main objective of that analysis was to clarify the nature of the
transition in $d=3$ from the paramagnetic phase to the ordered phase
which the authors concluded was most likely of weakly
first order both for $N=2,3$. It was further shown that, already within
the approximation given in Eq.~(\ref{eq:EffActDE}), the NPRG approach
reproduces the one-loop results from a $d=2+\epsilon$ expansion
of the NL$\sigma$M, the leading term of the usual $d=4-\epsilon$ 
expansion and
also  the leading term of the large $N$ expansion. They did
however not discuss in detail the physics in $d=2$ beyond the leading
terms which recovers the one-loop NL$\sigma$M result. This is the 
main objective of the present work.

We extend the previous truncations of the effective average action
in two ways. 
First, studies of the BKT
transition\cite{Berges02,Gersdorff01} 
have shown that it is important not to truncate in the
power of the fields, 
and we therefore include local terms
to arbitrary power in the invariant $\rho$. The reason for this
is that in $d=2$ all local terms are relevant since in $d=2$ the engineering
dimension of the fields vanishes, as measured relative to the 
Gaussian fixed point. Secondly, we extend the  terms present in 
Eq.~(\ref{eq:EffActDE}) to fully non-local ones which effectively includes 
terms to arbitrary order in the spatial derivatives. 
This
gives a more accurate approximation of the model than if one would only
keep
leading order derivatives and is also not too difficult to implement.
We therefore 
write $\Gamma$
as a sum of a local and a non-local part,
\begin{equation}
\Gamma_\Lambda[{\bd \Phi}_1,{\bd \Phi}_2]=
\Gamma_\Lambda^{\rm loc}[{\bd \Phi}_1,{\bd \Phi}_2]+\Gamma_\Lambda^{\rm
  nloc}[{\bd \Phi}_1,{\bd \Phi}_2] \,
\label{eq:GammaLocNLoc}
\end{equation}
where the local part is of the form
\begin{equation}
\Gamma_\Lambda^{\rm loc}[{\bd \Phi}_1,{\bd \Phi}_2]=\int_x
U_\Lambda(\rho,\tau) \, ,
\end{equation}
and $U_\Lambda$ is a function of the two invariants
$\rho$ and $\tau$.
  These two
invariants are in fact the only local $O(N)\times O(2)$ 
invariants  in the sense
that all higher order invariant local 
terms can be expressed by them.\cite{Delamotte04}

\subsection{Approximation for the local potential}
\label{subsec:local}
Ideally, one would like to solve the full local potential exactly,
which is numerically very difficult and which we therefore
did not pursue. 
We have instead tried two
different approaches, the first based on a field expansion 
of $U_\Lambda(\rho,\tau)$ up to
eighth order in the field. However, we found that the field
expansion to a given finite order does not work very well since the
higher order vertices become dominant and drive either $\mu_\Lambda$
or $\lambda_\Lambda$ to negative values which leads to a  breakdown of the
flow at still quite large values of $\Lambda$. This is discussed
 in Appendix \ref{app:fieldexpansion}. In the other, more successful,
approach, we approximate the local potential
as $U_\Lambda(\rho,\tau)\approx V_\Lambda(\tau) +W_\Lambda(\rho)$.
We then keep the full field dependence of $W_\Lambda(\rho)$, but
approximate $V_\Lambda(\tau)$ by its leading term
in a field expansion,
\begin{align}
  U_\Lambda(\rho,\tau)& \approx \mu_\Lambda^0 \tau/4 +W_\Lambda(\rho).
\label{eq:Local}
\end{align} 
This choice is based on the assumption that the $\rho$ dependence
of the local potential is more important than the $\tau$ dependence
since it controls the symmetry breaking expectation value $\kappa_\Lambda$. 
Within such an ansatz, one avoids the problems coming from the large
higher order terms which appear in a finite
order field expansion. Note that such a scheme could also be
carried out to a higher order in $\tau$, which, at least for smaller
powers of $\tau$, would be significantly less numerically demanding 
than keeping
the full local potential.
Here we limit the analysis
however to the approximation (\ref{eq:Local}). 

\subsection{Approximation of the non-local terms}
\label{subsec:nonlocal}
The non-local part contains terms up to quartic order in the fields and,
as a direct generalization of Eq.~(\ref{eq:EffActDE}),
is approximated as
\begin{align}
  \Gamma_\Lambda^{\rm nloc}[{\bd \Phi}_1,{\bd \Phi}_2]&=\frac{1}{2}\int_k 
z_\Lambda(k)
  \big[ {\bd \Phi}_{1,k} \cdot  {\bd \Phi}_{1,-k} +{\bd \Phi}_{2,k} \cdot
  {\bd \Phi}_{2,-k} \big] 
  \nonumber \\ &
  +\frac{1}{4}\int_{x ,x'}\lambda_\Lambda(x-x')
\big[ \rho_x/2 -\kappa_\Lambda \big]\big[ \rho_{x'}/2 -\kappa_\Lambda 
\big]   \nonumber \\ &+
\frac{1}{4}\int_{x,x'} \frac{\mu_\Lambda(x-x')}{2}{\rm Tr} {\cal A}_x {\cal
  A}_{x'} 
\nonumber \\
& - \frac{1}{8}\int_{x,x'} \omega_\Lambda(x-x') 
 \nonumber \\
& \, \, \, \, \,
\times \big[{\bd \Phi}_1(x) \cdot {\bd \Phi}_2(x')-
 {\bd \Phi}_2(x) \cdot {\bd \Phi}_1(x') \big]^2
\label{eq:Nonlocal}
\end{align}
where 
$\int_k=(2\pi)^{-d}\int d^d k$, 
 $\rho_x = {\bd \Phi}_{1,x}^2 +{\bd \Phi}_{2,x}^2 $ is the $x$-dependent
'density' which measures the local fluctuating moment of the
120$^\circ$ magnetization 
and ${\cal A}_x=  {^t}\Phi_x \Phi_{x} - \eins \rho_x/2$ are $x$-dependent
matrices. The $\mu$ part of the action can also be written as
\begin{align}
    \frac{1}{2}{\rm Tr} {\cal A}_x {\cal A}_{x'} = & ({\bd
    \Phi}_{1,x}^2-{\bd \Phi}_{2,x}^2) ({\bd
    \Phi}_{1,x'}^2-{\bd \Phi}_{2,x'}^2) /4 \nonumber \\ &
  +{\bd \Phi}_{1,x} \cdot {\bd \Phi}_{2,x} \,
  {\bd \Phi}_{1,x'} \cdot {\bd \Phi}_{2,x'} \, .
\end{align}
The coupling functions $z_\Lambda$, $\mu_\Lambda$, $\lambda_\Lambda$
and $\Omega_\Lambda$ are all defined to be completely non-local,
i.e. after a  Fourier transform they have a vanishing
contribution at momentum $k=0$. The local contributions are included
in $U_\Lambda$ and will be denoted by $\mu_\Lambda^0$ and $\lambda_\Lambda^0$.
For later convenience we also introduce the functions
\begin{subequations}
\begin{align}
\bar{\mu}_\Lambda(k)&=\mu_\Lambda(k)+\mu_\Lambda^0 \, , \\ 
\bar{\lambda}_\Lambda(k)&=\lambda_\Lambda(k)+\lambda_\Lambda^0 \, .
\end{align}
\label{eq:tildes}
\end{subequations}
The derivative terms present in the action (\ref{eq:EffActDE})
correspond to the approximation 
\begin{subequations}
\begin{align}
z_\Lambda(k)&= Z_\Lambda k^2 + {\cal O}(k^4) \, , 
\label{eq:zExp}
\\
\omega_\Lambda(k)&= \Omega_\Lambda k^2 + {\cal O}(k^4) \, ,
\label{eq:OmExp}
\end{align}
\end{subequations}
and $\mu_\Lambda(k)=\lambda_\Lambda(k)=0$. To keep also 
the leading order $k^2$
terms of $\mu_\Lambda(k)$ and $\lambda_\Lambda(k)$ 
is equivalent to introducing the derivative terms
\begin{align}
&\left( {\bd \Phi}_1 \cdot \partial_x {\bd \Phi}_1+ 
{\bd \Phi}_2 \cdot \partial_x {\bd \Phi}_2 \right)^2  \, ,  \nonumber \\
&\left( {\bd \Phi}_1 \cdot \partial_x {\bd \Phi}_1- 
{\bd \Phi}_2 \cdot \partial_x {\bd \Phi}_2 \right)^2
+\left( {\bd \Phi}_1 \cdot \partial_x {\bd \Phi}_2+ 
{\bd \Phi}_2 \cdot \partial_x {\bd \Phi}_1 \right)^2 \nonumber
\end{align}
in the action (\ref{eq:EffActDE}).

\subsection{Derivation of the flow equations}
\label{subsec:deriv}
The NPRG is based on an exact flow equation for the effective 
average action $\Gamma_\Lambda[\Phi]$,\cite{Wetterich93}
\begin{align}
\partial_\Lambda \Gamma_\Lambda[\Phi]= \frac{1}{2} \mbox{Tr} \left[
\partial_\Lambda R_\Lambda \left(\frac{\partial^2 
\Gamma_\Lambda}{\partial \Phi \partial \Phi} +R_\Lambda \right)^{-1} \right],
\label{eq:Wetterich}
\end{align}
where the trace is over momenta and internal indices. 
For notational brevity we omitted internal indices and momenta
in the field derivatives as well as in $R_\Lambda$ in
Eq.~(\ref{eq:Wetterich}). Note that the second order field derivative
of $\Gamma_\Lambda$ on the r.h.s. in Eq.~(\ref{eq:Wetterich}) is also
a functional of the field $\Phi$. If both sides of this
equation are expanded in the fields, one obtains flow equations
for the irreducible vertices.
The derivation of Eq.~(\ref{eq:Wetterich})
is based on an approach where the cutoff is introduced into
the model via a regulator $R_\Lambda$ which is added to
the bare two-point function. At the initial UV scale $\Lambda_0$,
the action is assumed to be the bare one, whereas
the full irreducible vertices
are obtained from Eq.~(\ref{eq:Wetterich}) when the flow
of $\Gamma_\Lambda$ is integrated from
$\Lambda=\Lambda_0$ down to $\Lambda=0$.
 While  Eq.~(\ref{eq:Wetterich}) is exact, it is 
almost always impossible to solve it exactly and approximation techniques
are required. The most common ones are either based on
an expansion of $\Gamma_\Lambda$ to a finite order in the fields
or an expansion in the derivatives, 
for reviews see 
e.g.~[\onlinecite{Berges02,Delamotte04,Kopietz10}]. Here, we will
use a combination of both, where we take into account both terms
which are not restricted to a finite order in the fields but also
non-local terms to arbitrarily order in the derivatives.\cite{Hasselmann12}
The regulator
$R_\Lambda$ 
 removes IR divergent terms arising
from modes with $k<\Lambda$ and for numerical stability we use
an analytic regulator. A standard choice\cite{Berges02} is
\begin{equation}
R_\Lambda(q)=Z_\Lambda \frac{q^2}{\exp(q^2/\Lambda^2)-1} \, .
\end{equation}
The flow equations are most easily derived in a basis where the
two-point functions 
are diagonal. Following Ref.~[\onlinecite{Delamotte04}]
we therefore first introduce the 
two $N$-component fields ${\bd
  \Phi}_{a}(x) ={\bd \varphi}_{a}(x) + {\bd \chi}_{a} $, where 
$ {\bd \chi}_{a} $ are the finite expectation values which we assume
to have the form
\begin{subequations}
\begin{align}
 {^t}{\bd \chi}_1& =(\kappa_\Lambda^{1/2},0,\dots,0) \, , \\
 {^t}{\bd \chi}_2& =(0,\kappa_\Lambda^{1/2},0\dots,0) \, .
\end{align}
\end{subequations}
Diagonalization of the two-point functions is achieved by a switch
to the basis $\tilde{\varphi}^\alpha_a$ (with $a=1,2$, and $\alpha=1\dots N$)
\begin{subequations}
  \begin{align}
    \tilde{\varphi}^1_1& =\frac{1}{\sqrt{2}} ( \varphi_1^1+\varphi_2^2) \, ,\\
    \tilde{\varphi}^1_2& =\frac{1}{\sqrt{2}} ( \varphi_1^1-\varphi_2^2) \, , \\
    \tilde{\varphi}^2_1& =\frac{1}{\sqrt{2}} ( \varphi_1^2+\varphi_2^1) \, , \\
    \tilde{\varphi}^2_2& =\frac{1}{\sqrt{2}} ( \varphi_1^2-\varphi_2^1)  \, ,
  \end{align}
\label{eq:rotated}
\end{subequations}
and $\tilde{\varphi}_a^\alpha=\varphi_a^\alpha$ for $\alpha>2$.
In the rotated basis
only one component has a finite expectation value, $\tilde{\chi}^a_\alpha=\sqrt{2 \kappa_\Lambda}
\delta_{a1}\delta_{\alpha 1}$. The 
two-point vertices are now diagonal in the $a,\alpha$ space and 
have the form 
\begin{align}
  \Gamma_{ab}^{\alpha \beta}(k) & = 
\frac{\partial^{(2)}}{\partial 
\tilde{\varphi}_{a}^{\alpha}(k)\tilde{\varphi}_{b}^{\beta}(-k)} \Gamma_\Lambda
\big|_{\varphi=0}
 \nonumber \\
&= \delta_{ab}\delta_{\alpha\beta}\Big\{
z_\Lambda(k)+\kappa_\Lambda 
\big[\delta_{a,1}\delta_{\alpha,1}\bar{\lambda}_\Lambda(k)+\eta_{a\alpha}
\bar{\mu}_\Lambda(k)\nonumber \\ &  \quad 
+\delta_{a,2}\delta_{\alpha,2} \omega_\Lambda(k)\big] \Big\} \, ,
\label{eq:gamma2}
\end{align}
where we introduced $\eta_{a\alpha}$ which has as 
nonzero entries only $\eta_{12}=\eta_{21}=1$. The functions 
$\bar{\mu}_\Lambda(k)$ and $\bar{\lambda}_\Lambda(k)$
are defined in Eqs.~(\ref{eq:tildes}).

The flow of the local potential is obtained by evaluating
Eq.~(\ref{eq:Wetterich}) for constant fields.\cite{Berges02}
The initial form of the local potential at $\Lambda=\Lambda_0$
coincides with the interaction term of the bare action and is given by
$U_{\Lambda_0}=\lambda_{\Lambda_0}^0 (\rho/2
 -\kappa_{\Lambda_0})^2/4+\mu_{\Lambda_0}^0\tau/4$. 
The flow of the local terms present in the two-parameter function
$U_\Lambda$ is greatly simplified if we employ the approximation
Eq.~(\ref{eq:Local}).
The flow of $W_\Lambda(\rho)$ is then found to be
\begin{align}
\partial_\Lambda W_\Lambda(\rho)&=\frac{1}{2} \int_k \partial_\Lambda
R_\Lambda(k) \sum_{a=1,2,\alpha=1\dots N} G_\Lambda^{a\alpha}(k,\rho)
\label{eq:flowW}
\end{align}
with 
\begin{subequations}
  \begin{align}
    G_\Lambda^{11}(k,\rho)&=\big[A_\Lambda(q,\rho)+\rho W^{\prime\prime}_\Lambda(\rho)+
    \rho \lambda_\Lambda(q)/2 \big]^{-1} \, , \\
    G_\Lambda^{12}(k,\rho)&=\big[A_\Lambda(q,\rho)+
    \rho \bar{\mu}_\Lambda(q)/2 \big]^{-1} \, , \\
    G_\Lambda^{22}(k,\rho)&=\big[A_\Lambda(q,\rho)+
    \rho \Omega_\Lambda(q)/2 \big]^{-1} \, .
  \end{align}
\end{subequations}
The remaining functions are $G_\Lambda^{21}(k,\rho)=G_\Lambda^{12}(k,\rho)$
and there are $2(N-2)$ modes of the form $G_\Lambda^{a\alpha}=A_\Lambda(q,\rho)^{-1}$ 
for $\alpha>2$ with
\begin{equation}
A_\Lambda(q,\rho)=R_\Lambda(q)+z_\Lambda(q)+W^\prime_\Lambda (\rho) \, . 
\end{equation}
The flow of
$\kappa_\Lambda$ is obtained from the requirement that
$(d/d\Lambda) W^\prime(\rho=2\kappa_\Lambda)$=0, i.e. that $\kappa_\Lambda$ 
is for all $\Lambda$ the position of the minimum of $W_\Lambda$. \cite{Berges02}

To solve the flow equations numerically, we need  to
have an accurate resolution of the local potential
around the flowing minimum $\kappa_\Lambda$. 
 Since for $d=2$ this minimum vanishes for $N\geq 3$
at some finite $\Lambda^*$, reflecting the finite correlation
length, we need to rescale the 
local potential. This is achieved by writing
\begin{equation}
W_\Lambda(\rho)=\kappa^2_\Lambda w_\Lambda(y=\rho/\kappa_\Lambda)
\end{equation}
so that the rescaled potential $w_\Lambda(y)$ always has
its minimum at $y=2$. Choosing a linear grid
for $y$ proved then sufficient to obtain converged
and stable flows. At low temperatures, $w_\Lambda(y)$ rapidly
approaches a convex form and becomes essentially flat for $y<2$.

To derive the flow equations of the non-local terms in $\Gamma_\Lambda$,
i.e. of the functions $\mu_\Lambda(k), \lambda_\Lambda(k)$ and $\Omega_\Lambda(k)$,
we  invoke a field expansion.\cite{Berges02,Kopietz10}
We need the vertices up to fourth order in an expansion in 
${\tilde\varphi}_a^\alpha$, they 
can be found in Appendix \ref{app:vertices}. From the explicit form of all
vertex functions up to the four point vertex, 
we can determine the flows of the non-local
coupling functions directly from the standard flows of the
two-particle vertices using Eq.~(\ref{eq:gamma2}), this
is discussed in detail in Refs.~[\onlinecite{Hasselmann12,Braghin10}].
The flow of the self-energy $\Gamma_{11}^{22}(k=0)$ 
also yields 
the flow of the local coupling constant $\mu_\Lambda^0$.
We emphasize that the obtained flow equations are uniquely determined by
the effective average action specified through 
Eqs.~(\ref{eq:GammaLocNLoc}-\ref{eq:Nonlocal}). The flow equations are
rather lengthy and not very illuminating, and we therefore
do not present them here. 
For an alternative
approach to include the momentum dependence of vertices,
which is not based on a truncation of the effective average action
but on an approximation at the level of a field expansion in
presence of a background field,
see Refs.~[\onlinecite{Benitez09}].

\begin{figure}[h]
\includegraphics[width=8.5cm,clip,angle=0]{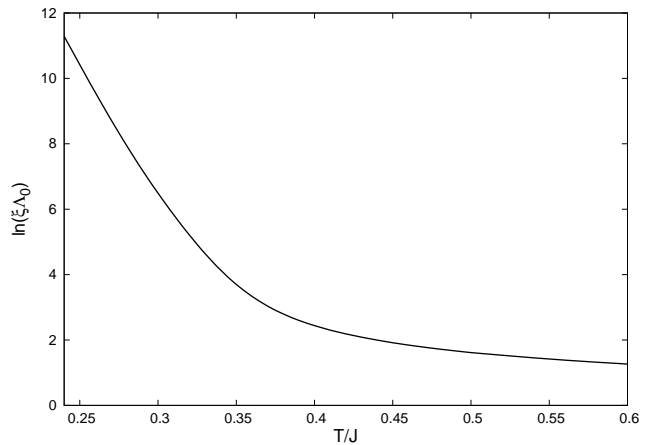}
\caption{The correlation length as a function of temperature has
an exponential dependence on the temperature but a much weaker
temperature dependence at larger temperature, with a crossover
temperature of
$T\approx 0.35 J$ separating the two regimes.}
\label{fig:xi}
\end{figure}

\begin{figure}[h]
\includegraphics[width=8.5cm,clip,angle=0]{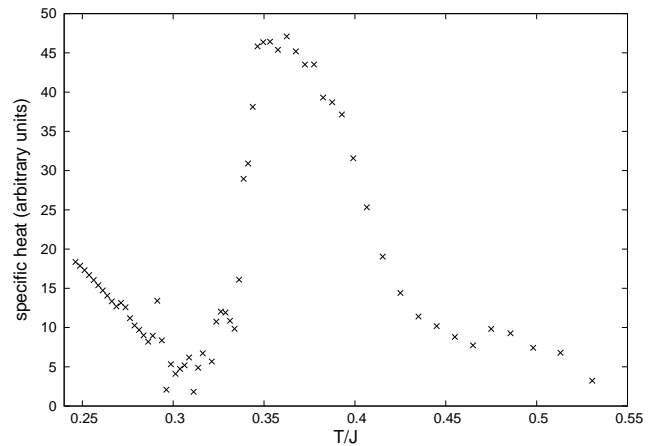}
\caption{The specific heat as a function of temperature. It shows
a well defined broad peak in the same temperature range where
the correlation length $\xi(T)$ has the crossover.}
\label{fig:specHeat}
\end{figure}

\section{Results and discussion}
\label{sec:results}
We have integrated the NPRG flow equations corresponding to
Eqs.~(\ref{eq:GammaLocNLoc}-\ref{eq:Nonlocal})
for different temperatures,
ranging from $T=0.6J$ down to $T=0.24 J$. The lower the temperature,
the smaller the logarithmic step size $\delta \ell$ 
(with $\ell=-\ln \Lambda/\Lambda_0$)
had to be chosen in the partial differential equation solver routine. 
If $\delta \ell$ is chosen too large in comparison with the discretization
$\delta y$ in the representation of the local potential $w_\Lambda(y)$
small oscillations in the derivatives of $w_\Lambda(y)$ appear which
quickly grow and lead to numerical instabilities. Thus, we had
to choose rather small step sizes at low temperatures, down to
$\delta \ell \simeq 4\times 10^{-5}$ for $T=0.24 J$, and we 
could not reach arbitrarily low temperatures since at $T\leq 0.24J$
it takes already more than a week
to calculate the flow for a given temperature (using a single core of the CPU).
 We also note that
an adaptive step solver turned out to be problematic since it
generally cannot cope well with the instabilities which arise
at larger step sizes.

We will be interested in the spin-correlation length $\xi$ which
characterizes the decay of the spin correlation function
$\big< {\bm S}_{{x}_i} \cdot {\bm S}_{{x}_j}\big>$ where
${x}_i$ and ${x}_j$ belong to the same sublattice of the 120$^\circ$
order. 
The relations (\ref{eq:spinmicro})
imply that 
$\big<{\bm \Phi}_{x} \cdot {\bm \Phi}_y \big>$
decays with the same correlation length as  
$\big< {\bm S}_{{x}_i} \cdot {\bm S}_{{x}_j}\big>$.
We can thus extract the spin correlation
length directly from our NPRG analysis as 
the scale $\Lambda^*$ where 
the order parameter $\kappa_{\Lambda^*}$ 
vanishes.
In Fig.~\ref{fig:xi} we show results for  $\xi$
as a function of temperature.  At low
temperatures the correlation length grows exponentially and follows
a $\xi \approx \exp B J/T$ behavior. 
We compare this with the correlation length of the  NL$\sigma$M applied
to
the HAFT, which, at two loop order, has the form \cite{Azaria92} 
\begin{equation}
  \xi_{\rm HAFT}/a\approx C_\xi \sqrt{T/J} \exp (6.9943 J/T) \, ,
\end{equation}
with an undetermined constant $C_\xi$. The exponent $B=6.9943$
is the same also in the one-loop approximation.
Our NPRG analysis yields the slightly smaller value $B\approx 6$. 
As we discuss in more detail below, the NPRG 
flow deviates
from the NL$\sigma$M flow already at moderate temperatures
and also at small spatial scales, although they
do coincide at small temperatures and large scales.
Since the NL$\sigma$M prediction for $\xi(T)$ is based
on the integration of the flow equation starting from
the lattice scale,  a small deviation of the 
correlation length exponent $B$ is not unexpected.

What can be clearly observed is
a pronounced crossover at around $T_{\rm cross}\approx 0.35 J$, from the
low temperature exponential temperature dependence to a 
much more modest decay of $\xi$ at larger temperatures. 
This
crossover happens in a relatively narrow temperature range,
yet $\xi(T)$ is smooth and continuous,
with no sign of an underlying thermodynamic singularity. This
crossover is similar to the sharp increase of the 
correlation length seen in MC simulations 
for temperatures $T\lesssim 0.3J$.\cite{Wintel95,Southern93}
The finite size limitations in MC combined with the exponential growth
of the correlation length make it however difficult to obtain converged 
results for $\xi(T)$ from MC and it was not
clear if $\xi(T)$ or its temperature derivative would be smooth in the
thermodynamic limit.
A recent theory\cite{Kawamura10} has proposed that the spin correlation length
is a convolution of two correlation lengths,
$\xi=\xi_v \xi_{sw}/(\xi_v+\xi_{sw})$. 
The vortex correlation length $\xi_v$ is assumed to
diverge at a finite temperature $T_c$ while the  spin-wave correlation length
$\xi_{sw}$
remains finite for all $T>0$. The resulting form of the magnetic correlation
length would have a non-monotonic function $d \ln(\xi\Lambda_0)/dT$
with a maximum near $T_{c}$. Our NPRG
analysis does not show such a behavior. We emphasize that this crossover
cannot be obtained within a NL$\sigma$M approach and neither in a finite
order field expansion, as discussed in Appendix \ref{app:fieldexpansion}.

The flow of the free energy can also readily be obtained from our analysis,
it corresponds to the flow of 
$\Gamma_\Lambda$ evaluated at $\rho=2 \kappa_\Lambda$ and $\tau=0$. Thus,
the flow of the free energy follows from Eq.~(\ref{eq:flowW}) with
$\rho=2\kappa$.
As a result of keeping the nonlocal coupling functions
$\mu_\Lambda(k)$, $\lambda_\Lambda(k)$ and $\Omega_\Lambda(k)$ in
our analysis, the thus evaluated free energy is sensitive to 
a broad range of energy scales beyond the IR limit.
Since the NPRG breaks down at some finite scale $\Lambda^*$ where
$\kappa_{\Lambda^*}=0$, we cannot follow the free energy flow 
down to $\Lambda=0$. To extract the contributions to the free energy
coming from $0<\Lambda<\Lambda^*$, we took
advantage of the fact that all propagators
are gapped in this regime because of the finite correlation length $\xi$ and thus
no IR divergences are present. 
We therefore
approximated the propagators in this regime
simply by introducing a finite correlation length $\xi^{-1}=2\pi \Lambda^*$
into the self
energies and by replacing all flow parameters by their values
at $\Lambda=\Lambda^*$. We note that the region $\Lambda<\Lambda^*$
contributes only a very small fraction to the total free energy at low
temperatures which has no noticeable effect on the shape of the
specific heat in the temperature range considered here.
From the thus obtained free energy we calculate
the specific heat $C=-T (\partial^2 f /\partial T^2)$, which 
required some local smoothing of the $f(T)$ data to avoid noise
in $C(T)$. Our result for $C(T)$ is plotted in  Fig.~\ref{fig:specHeat} 
and shows  a well defined
but relatively broad peak, again rather similar to what is obtained from
MC simulations.\cite{Kawamura10} While the specific heat typically 
shows a singularity near a second order phase transition with
a divergent correlation length, the broad peak observed here 
is a consequence of the rapid crossover of the spin correlation
length in that temperature regime rather
than a true divergence.

\begin{figure}[h]
\begin{minipage}[b]{8.5cm}
\includegraphics[width=4.0cm,clip,angle=0]{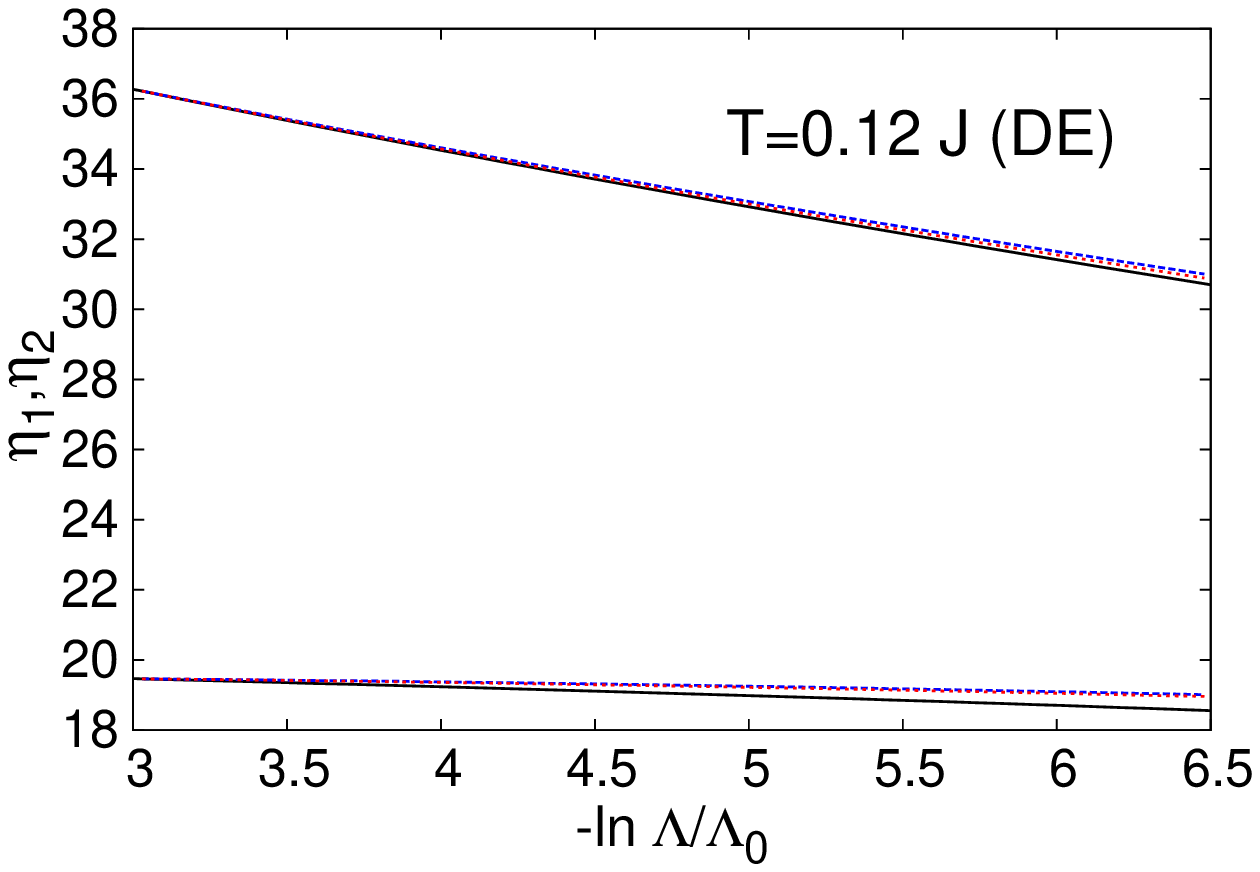}
\includegraphics[width=4.0cm,clip,angle=0]{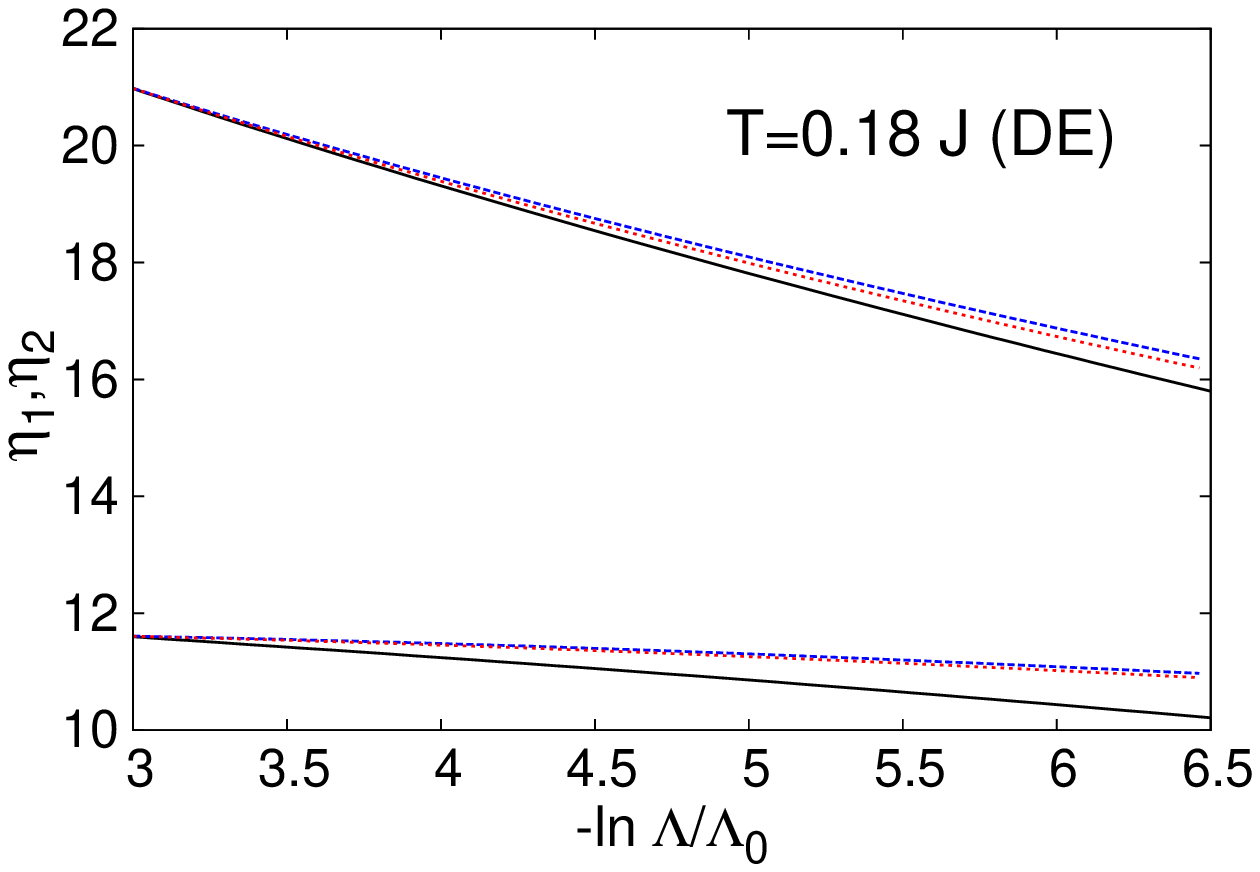}
\\
\includegraphics[width=4.0cm,clip,angle=0]{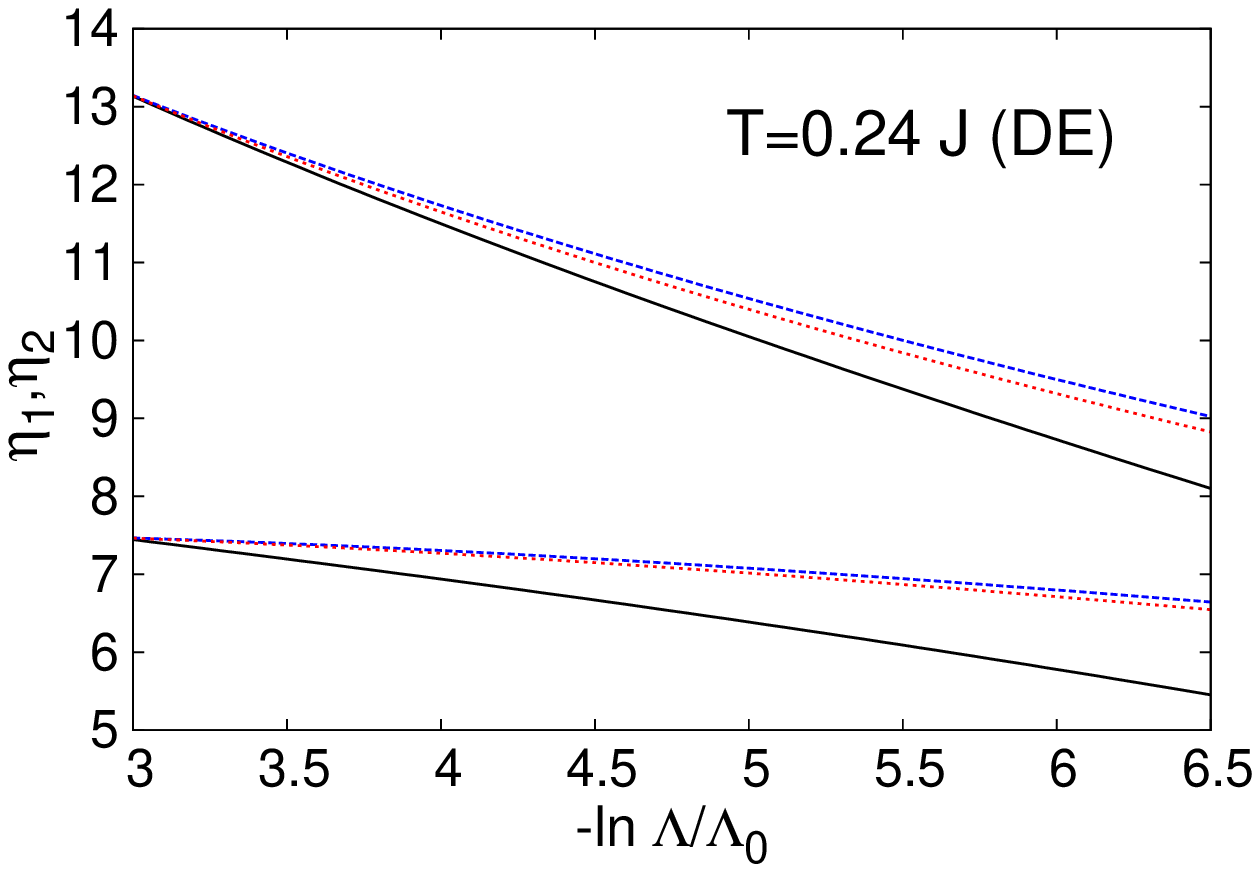}
\includegraphics[width=4.0cm,clip,angle=0]{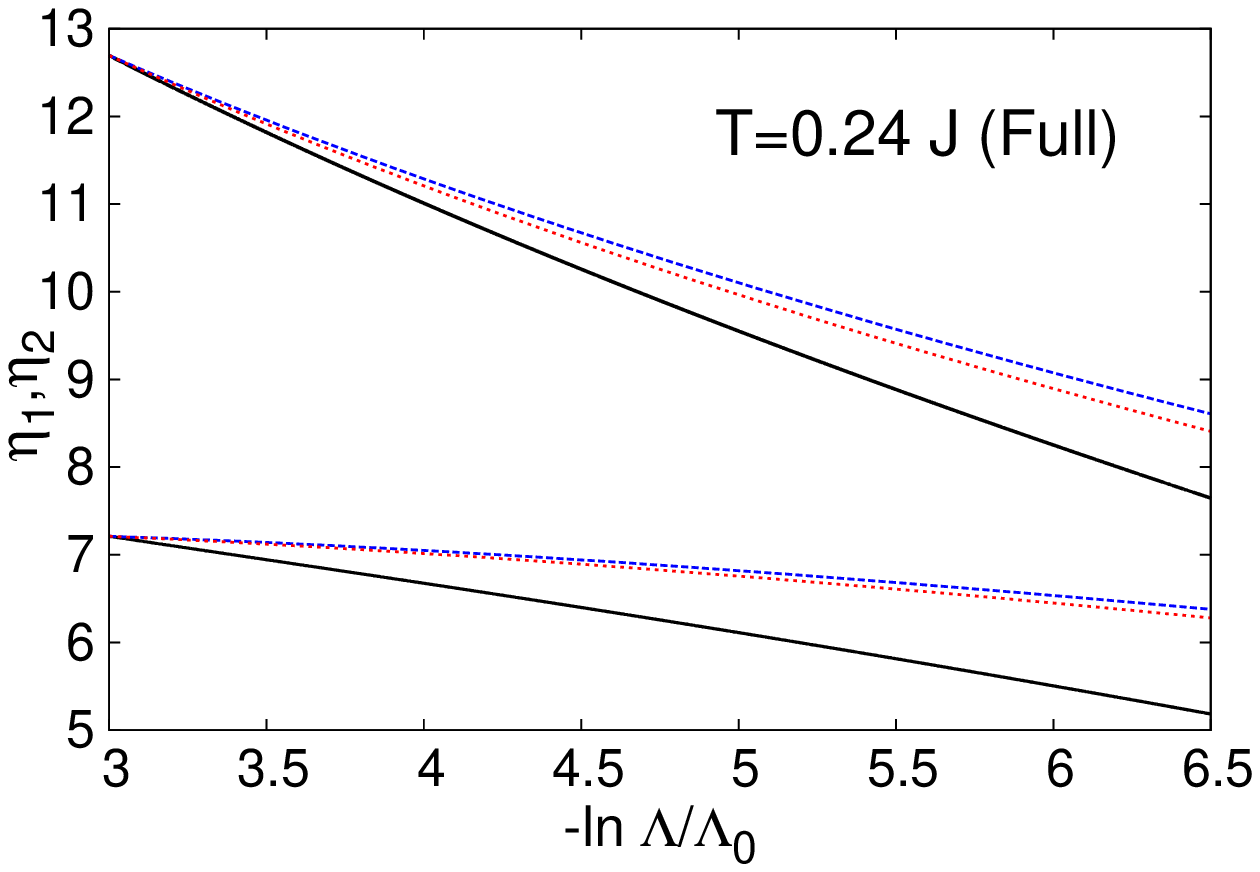}\\
\includegraphics[width=4.0cm,clip,angle=0]{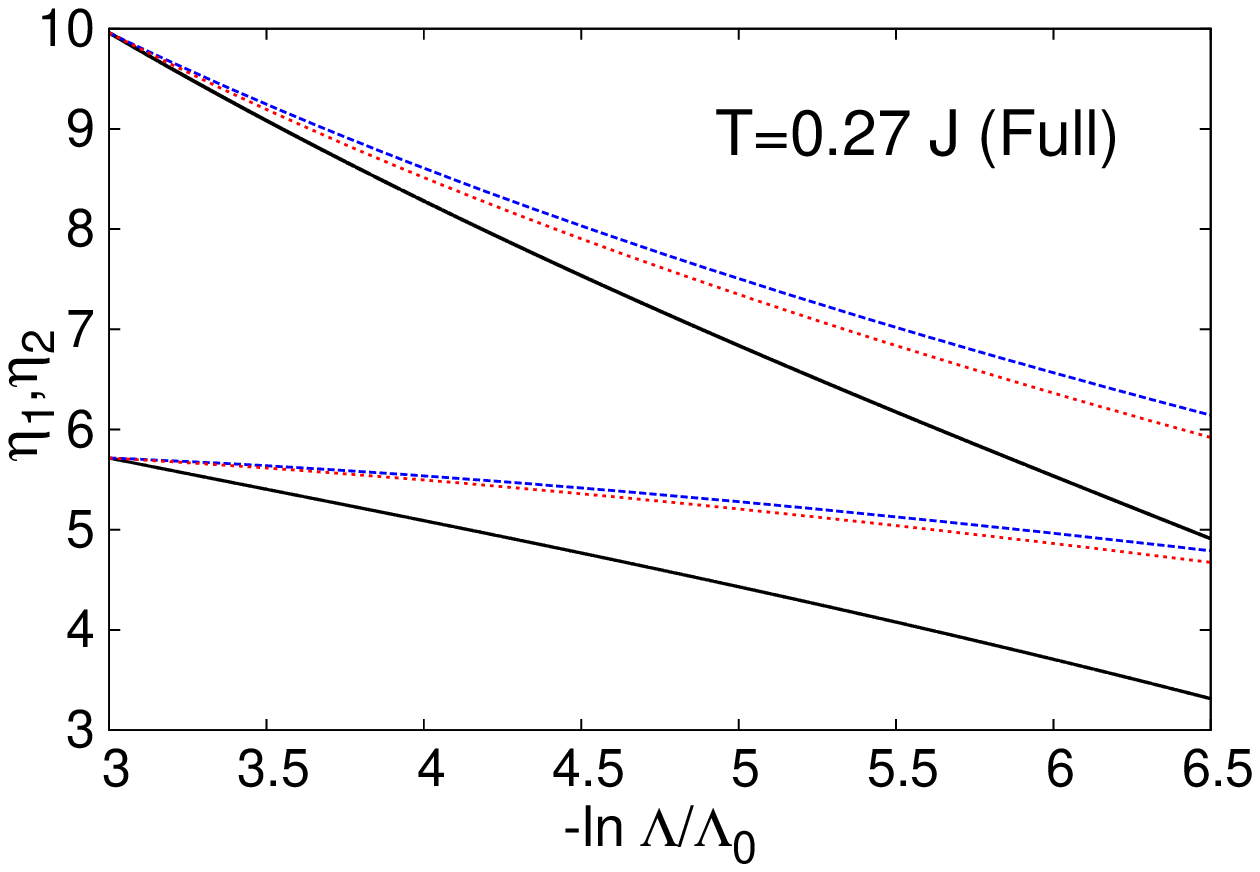}
\includegraphics[width=4.0cm,clip,angle=0]{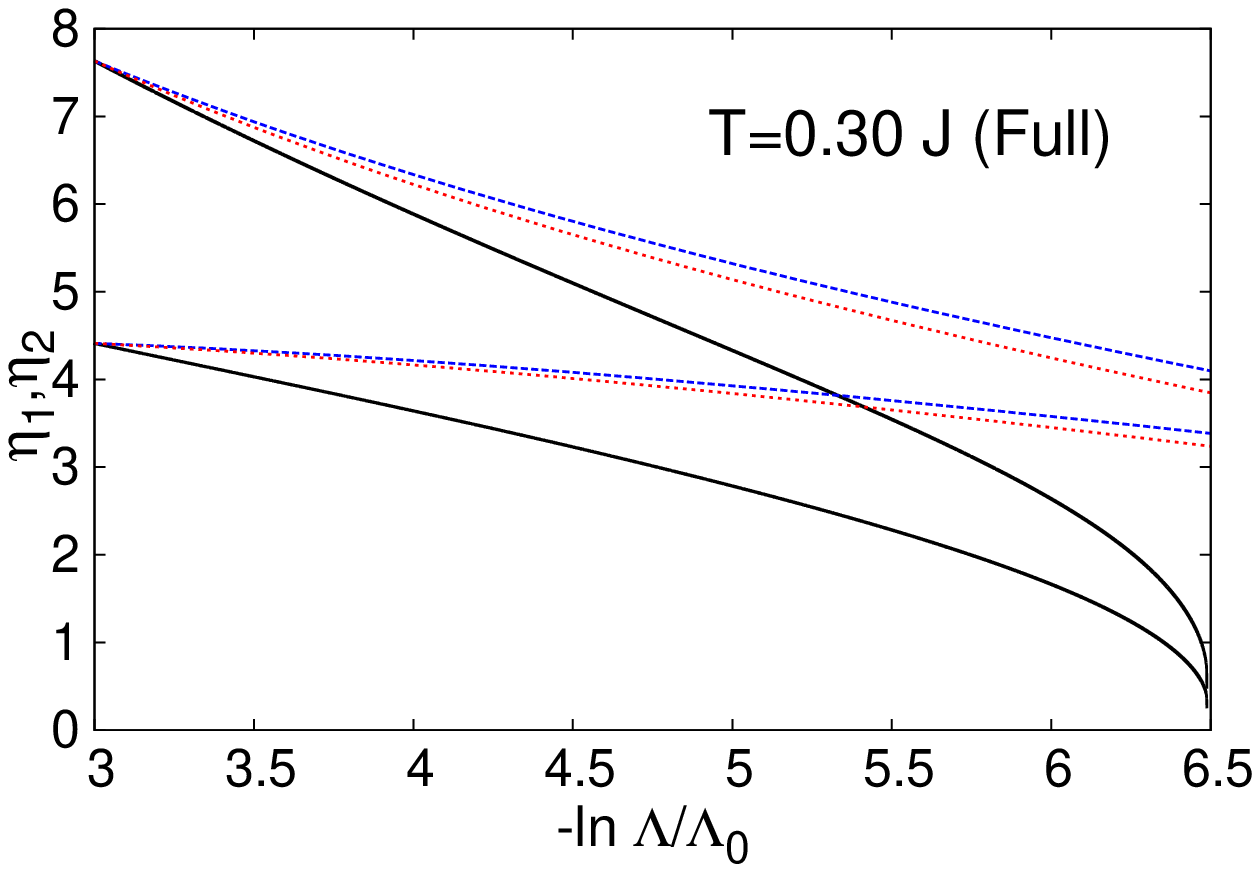}\
\caption[]{(Color online) Flow of the parameters $\eta_1$, $\eta_2$
(with $\eta_2>\eta_1$), from NPRG calculations
(solid black lines) and from perturbative one-loop (dotted blue lines) and 
perturbative two-loop (dashed red lines) RG calculations for the
NL$\sigma$M. For the NL$\sigma$M $\eta_{1,2}$ are the spin-stiffnesses
whereas for the NPRG we extracted $\eta_{1,2}$ via 
Eqs.~(\ref{eq:eta1},\ref{eq:eta2}).
For low temperatures $T\leq 0.24 J$ the NPRG flows
are those of the derivative expansion (DE), see  (\ref{eq:EffActDE}), 
whereas for $T\geq 0.24J$ we
show the flow calculated in the full approximation 
Eqs.~(\ref{eq:GammaLocNLoc}-\ref{eq:Nonlocal}). 
The one- and two- loop NL$\sigma$M results are always very similar and strongly
overlap in the plots for $T\leq 24$. For $T=0.12$ all approximations 
strongly overlap.
}
\label{fig:eta1eta2}
\end{minipage}
\end{figure}

The behavior both of the correlation length and the specific heat
thus suggest that there is no true phase transition at
$T_{\rm cross}$ but rather a crossover from a purely spin-wave dominated
regime (the NL$\sigma$M regime) to a high temperature
regime where defects and massive excitations
play an important role. 
This picture is also supported by comparing
our NPRG flow to the one obtained from the NL$\sigma$M. In
Fig.~\ref{fig:eta1eta2} we show the flow of the two parameters
$\eta_1$ and $\eta_2$ from the NPRG and for the NL$\sigma$M,
at different temperatures. Since
we have data from the full NPRG flow 
 only down to $T=0.24J$, we also show data obtained from 
the derivative expansion of $\Gamma_\Lambda$, using only the parameters
which enter in Eq.~(\ref{eq:EffActDE}). In the limit of 
large masses $\lambda_\Lambda^0$ and $\mu_\Lambda^0$
the NPRG flow equations in the derivative expansion approximation
reduce to  the one-loop NL$\sigma$M flow.\cite{Delamotte04}
This is clearly seen at $T=0.12J$. However, the large mass limit of the 
flow equations is reached only very slowly and at moderately
small temperatures finite mass corrections are visible. 
Already at $T=0.18J$ one sees 
deviations from both the one-loop and two-loop NL$\sigma$M flow, 
which become quite substantial at $T=0.24J$. At this temperature
we have calculated the NPRG flow both in the full approximation,
corresponding to Eqs.~(\ref{eq:Local},\ref{eq:Nonlocal}), as well
as in the derivative approximation (\ref{eq:EffActDE}). Both
NPRG flows show similar deviations from the NL$\sigma$M results.
The deviations grow at even larger temperatures when compared
to the full NPRG flow. At $T=0.30J$, $\eta_1$ and $\eta_2$ vanish
at a $\Lambda$ scale where the NL$\sigma$M results still predict
sizeable finite stiffnesses.

\begin{figure}[h]
\begin{minipage}[b]{8.5cm}
\includegraphics[width=8.0cm,clip,angle=0]{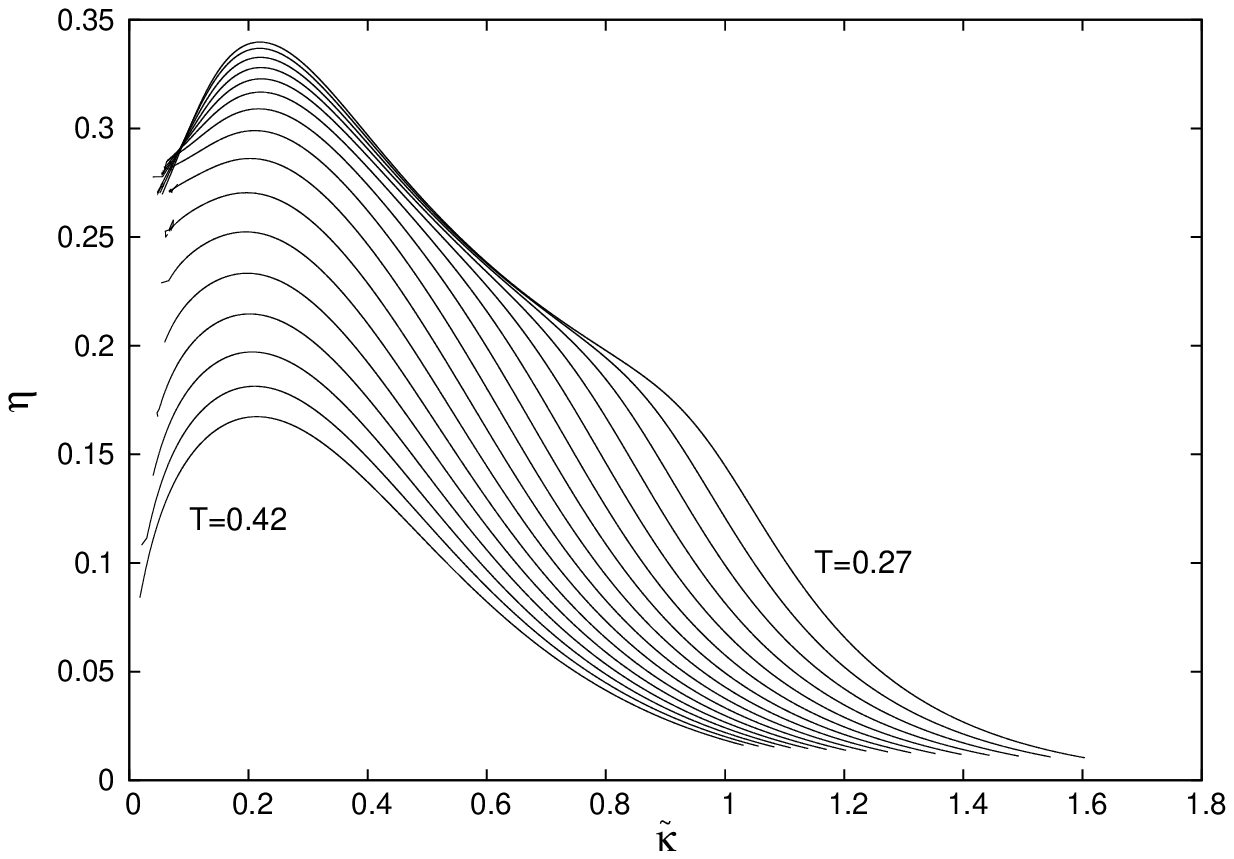}\\
\includegraphics[width=8.0cm,clip,angle=0]{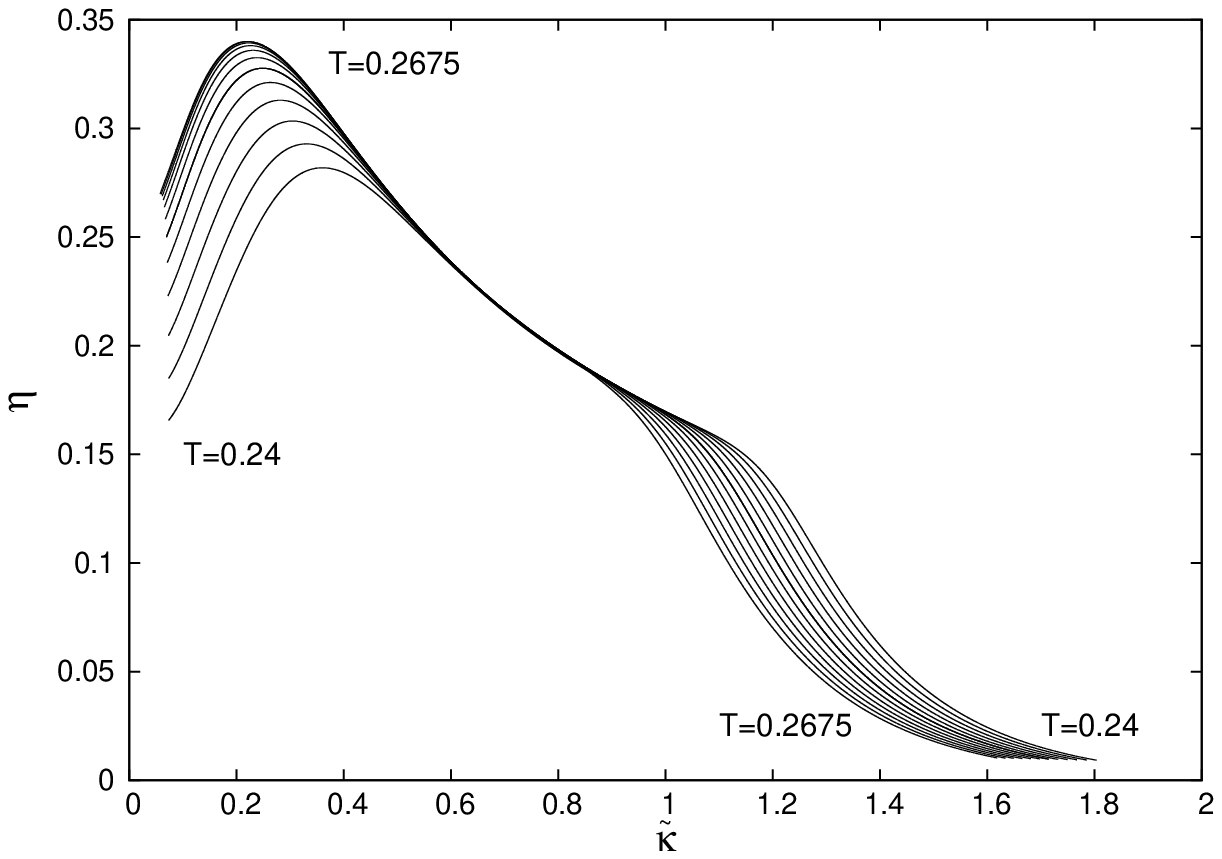}
\caption[]{Anomalous exponent $\eta$ vs. the rescaled order parameter 
$\tilde{\kappa}$. Upper curve shows results for large temperatures,
the lower curve at smaller temperatures.}
\label{fig:etavskappa}
\end{minipage}
\end{figure}

The crossover at $T_{\rm cross}$ has often
been argued to be caused by
unbound
$Z_2$ defects which start to proliferate at $T_{\rm cross}$.\cite{Kawamura10,Southern95} 
MC simulations have found evidence of a vortex
unbinding at this temperature. Southern and Xu \cite{Southern95}
have extracted a vorticity modulus from their MC data which was
shown to vanish near $T_{\rm cross}$ which was interpreted as
an unbinding of vortices. 
While the spin stiffness is always zero
for any $T>0$, in finite sized systems the spin stiffness 
vanishes only at sufficiently large $T$. For system sizes comparable
to those of Ref.~[\onlinecite{Southern95}] the spin stiffness
vanishes at roughly the same $T$ as the vorticity modulus.\cite{Wintel95}
The vortex unbinding at low temperatures is prevented by  
a logarithmic interaction among the vortices which is however
only present for length scales smaller than the correlation length.\cite{Wintel94}
To gain further insight
into the physics behind the crossover we plot the flowing
anomalous dimension $\eta$.
For a true $2^{nd}$ order phase transition $\xi \to \infty$ and $\eta$ would 
become a critical exponent. It
characterizes the spin-spin correlation function 
at criticality 
which behaves for 
$k\to 0$ as $1/k^{2-\eta}$. Here, for any finite $T$ the spin correlation
length is finite and $\eta$ does not reach a constant for $k\to 0$.
Yet, for low temperatures, $\eta$ changes only very modestly
for momenta   $k<1/\xi$ 
and both $\xi$ and $2\pi/k$ much larger
than the microscopic lattice spacing. 
The scale-dependent anomalous dimension $\eta$ is
defined through
\begin{equation}
\eta=\Lambda \partial_\Lambda \ln Z_\Lambda \, .
\end{equation}
We plot it 
as a function of the 
rescaled
order parameter $\tilde{\kappa}$, defined in Eq.~(\ref{eq:kapparescaled}), 
for different temperatures in
Fig.~\ref{fig:etavskappa}. For the XY-model, these plots
show a characteristic flow  which,
for temperatures below the critical one, quickly reaches a
line of $\eta(\tilde{\kappa})$
where
the flow of $\tilde{\kappa}$ essentially stops.\cite{Gersdorff01}
This line signifies thus a line of fixed points where the
anomalous dimension reaches a finite value for $\Lambda\to 0$.
The line of fixed points terminates around a value $\eta=0.287$,
beyond which the flow is away from the line of fixed points.
What we find in the present model is in some ways similar
to the XY flow, with
however important differences. At low temperatures
we do find a common curve 
$\eta(\tilde{\kappa})$ where all flows are attracted to.
However, while the flow along this line is slower
that the initial approach to that line, the flow never 
stops but remains sizeable, in accordance with
the asymptotic freedom of the model. Thus, one never actually
reaches a fixed point and no transition or critical
behavior can be associated with the common curve.

The maximal value of $\eta_{\rm max} \approx 0.34$, is close
to the value $\eta_{\rm max}\approx 0.33$ found in the XY model 
(the line of fixed point ends in the XY model at a smaller
value $\approx 0.287$, rather close to the exact value $1/4$).
This may be seen as an indication that indeed some enhanced
stability against defect unbinding exist along the line
$\eta(\tilde{\kappa})$.
In this interpretation, for temperatures lower than 
roughly $T\approx 0.27J$, a large part of the flow is 
along the common curve $\eta(\tilde{\kappa})$ and 
only at sufficiently small $\Lambda$ it deviates, owing
to the vanishing of the order parameter and the appearance
of a finite correlation length. 
While defects would certainly be
present at scales larger than the correlation length, 
the fact that
the maximal
value of $\eta$ systematically 
decreases with lowering the temperature below $T=0.27J$
indicates a stability against vortex unbinding.
Thus, the correlation length in this regime
is limited by the asymptotic freedom of the model rather
than an unbinding of vortices.
The stability to vortices (for scales smaller than the
correlation length) is then similar to the XY model where
along the critical line $T<T_c$ the anomalous dimension reaches
a fixed point whose value is proportional to the temperature. 
We also find that over a large momentum range the spin correlation decay is
algebraic with anomalous exponents and only at large distances 
the finite correlation
length induces an exponential decay.

At larger temperatures, see the upper plot in Fig.~\ref{fig:etavskappa},
the flow starts to deviate slowly from the common curve $\eta(\tilde{\kappa})$
of the low temperature regime. At around $T\approx 0.30J$
it never reaches it  and moves further
away from it the higher the temperature. This indicates that
at these temperatures the regime where gapless excitations
dominate the flow is never reached and  massive and/or
topological excitations become ubiquitous. 
For $T=0.35$ the
maximum is approximately $\eta=0.28$, similar to the NPRG
estimate of the critical temperature of the XY model. 
All this is consistent with a correlation length which
is, at high temperatures, primarily determined by 
a vortex unbinding. That  the correlation lengths
actually decreases more slowly at higher temperatures
where vortices are abundant, as is also observed
in MC simulation,\cite{Wintel95,Southern93} 
can be understood on the
grounds that the flow ceases to be controlled by the
low temperature NL$\sigma$M model and its strong,
asymptotic freedom dominated, temperature dependence
of the correlation length.

Thus, we see in the flow of $\eta$
support for the scenario of vortex unbinding somewhere
in the temperature interval $0.3J-0.35J$, where also the
crossover in the correlation length dependence on temperature is
observed. 
In the NPRG we have however no direct access to
vortex degrees of freedom, so that we can only say that
our results are consistent with a vortex unbinding scenario.

\section{Summary}
\label{sec:summary}

We have analyzed $2d$ frustrated Heisenberg magnets within a non-perturbative
RG framework, using initial values for the flow as appropriate for the
antiferromagnetic Heisenberg model on the triangular lattice. We follow
the general NPRG approach for frustrated magnets as developed in 
Refs.~[\onlinecite{Tissier00,Delamotte04}], which allows to recover
the NL$\sigma$M flow equations at sufficiently low temperatures.
We extend this analysis in two
ways: instead of expanding the action to a given order in 
a field expansion
we keep the full local dependence of the effective action on the
invariant $\rho$ which is the local fluctuating magnitude of the ordered moments
of the magnet. Further, we replace the coupling parameters of the standard
Landau-Ginzburg action with non-local coupling functions.

The primary goal of our analysis is to clarify the nature of the finite
temperature crossover in the correlation length dependence
of the MC simulations of the HAFT at around a temperature $T\approx 0.3 J$
and to investigate a possible role of topological defects.
Our analysis reproduces the key feature of the MC simulations.
As expected, at low-temperatures we recover the flow of the NL$\sigma$M
which was shown to be in accordance with MC simulations in 
Refs.~[\onlinecite{Southern93,Wintel95}]. The NPRG flow deviates
further and further from the NL$\sigma$M predictions upon increasing
the temperature and 
we find 
a crossover of  the temperature dependence of the correlation length
at around $T_{\rm cross}\approx 0.35J$. Although this temperature 
is slightly larger than the one
observed in MC, which is not surprising in view of the approximations
inherent in the mapping of the lattice model into a continuum theory,
our NPRG approach does capture the crossover qualitatively. 
The specific heat, which shows a broad
peak around the crossover, is also 
in qualitative agreement with MC simulations.\cite{Kawamura10}
As discussed in Sec.~\ref{sec:results}
closer inspection of the flow of the anomalous dimension shows that
for temperatures  slightly lower than $T_{\rm cross}$ the flows collapse over
a wide range of scales
on a common curve $\eta(\tilde{\kappa})$, where $\tilde{\kappa}$
is the rescaled local order parameter. In this regime the (large) correlation length 
arises from the asymptotic freedom of the model and
thus from the geometry of the order parameter space. In contrast,
at around temperatures $0.3J-0.35J$, the flow starts to
deviate from the common curve $\eta(\tilde{\kappa})$. 
The maximal anomalous dimension is reached around this temperature 
and is similar to that of the XY model at the
vortex unbinding transition. While a topological origin of this
behavior is plausible,
we find no indication of a finite temperature fixed point and
all our results are instead consistent with a crossover.
Physically, this is not completely unexpected since  a true 
phase transition 
would usually require a logarithmic interaction among the vortices,
as it occurs within the BKT scenario.
In view of the finite correlation length the logarithmic interaction
is cut off at large distances, which would result in a crossover rather
than a phase transition, and there would be no diverging length scale
associated with the crossover.
How can this be reconciled with
the presence of two phase transitions which are clearly observed
in MC simulations at small but finite fields?\cite{Mike11,Shannon11} 
A likely scenario is that for vanishing fields the two critical points
meet and merge into a crossover point instead. This is consistent with
the vanishing of the order parameters of both low temperature phases
in the zero field limit.\cite{Shannon11}
Further analysis
of the low field regime with the NPRG  would certainly be desirable.

We thank Federico Benitez, Nic Shannon, Dominique Mouhanna and
Lorenz Bartosch for discussions and suggestions.
N.~H. acknowledges support from the DFG research group FOR 723.

\bigskip

\bigskip
\bigskip

\appendix

\section{Form of higher order vertices}
\label{app:vertices}
Here we give the expressions for
 the symmetrized
vertices which are required to derive the flow equations
of the non-local coupling functions. Besides the two-point vertex, given
in Eq.~(\ref{eq:gamma2}), these are the three- and four point vertices. 
The three point vertex, in the basis defined in Eqs.~(\ref{eq:rotated}), is
\begin{widetext}
\begin{align}
  \Gamma_{a_1a_2a_3}^{\alpha_1 \alpha_2 \alpha_3}({\bm k}_1,{\bm k}_2,{\bm k}_3)&=\sqrt{\kappa/2} \Big\{
  \big[ 
\delta_{a_11}\delta_{\alpha_12} \big(\delta_{\alpha_2\alpha_3}^> \eta_{a_2
  a_3} +\eta_{\alpha_2 \alpha_3}\delta_{a_2 a_3} \xi_{a_2} \big)
 + \delta_{a_1 2} \delta_{\alpha_11} \big( \delta_{\alpha_2\alpha_3}^>
 \delta_{a_2 a_3} \xi_{a_2}+\delta_{\alpha_2\alpha_3}^< \eta_{a_2 a_3} \big)
\big] 
\mu(k_1) 
\nonumber \\
  & \quad
  +\delta_{a_1 1}\delta_{\alpha_1
    1}
\delta_{a_2 a_3} \delta_{\alpha_2 \alpha_3} [ \lambda(k_1) 
+32\kappa U_\Lambda^{\prime\prime\prime}(\rho=2\kappa) \delta_{a_21} \delta_{\alpha_21} ]
\nonumber \\
  & \quad
  +\delta_{a_1 2} \delta_{\alpha_1 2} 
  \epsilon_{a_2 a_3} [\delta_{\alpha_2 \alpha_3}-\Theta(\alpha_2\leq
  2)\Theta(\alpha_3\leq 2)] 
\big[ \Omega(k_2)-\Omega(k_3) \big]
  + (1\leftrightarrow 2) + (1\leftrightarrow 3) \Big\} \, .
\end{align}
\end{widetext}
where $\delta_{\alpha\beta}^<=\delta_{\alpha \beta} \Theta(\alpha\leq 2)$,
and  $\delta_{\alpha_\beta}^> = \delta_{\alpha \beta} \Theta(\alpha> 2)$.
We further defined 
the vector
$\xi_a= (1,-1)^t$ and $\epsilon_{ab}=-\epsilon_{ba}$ is the antisymmetric
tensor with $\epsilon_{12}=1$. The tensor $\eta_{a\alpha}$ has nonzero
entries only for $\eta_{12}=\eta_{21}=1$.
The notation
$(1\leftrightarrow 2)$ is short for ($k_1, a_1,\alpha_1 \leftrightarrow k_2, a_2,\alpha_2$).
The four point vertex in the basis defined in Eqs.~(\ref{eq:rotated}) is
\begin{widetext}
\begin{align}
  \Gamma_{a_1\dots a_4}^{\alpha_1 \dots \alpha_4}({\bm k}_1,\dots,{\bm k}_4)&=
  \frac{1}{2} \Big\{ \Big[ \big( \eta_{a_1a_2}\delta_{\alpha_1 \alpha_2}^>
+\delta_{a_1a_2} \xi_{a_1} \eta_{\alpha_1 \alpha_2} \big) 
\big(\eta_{a_3a_4}\delta_{\alpha_3 \alpha_4}^>
+\delta_{a_3a_4} \xi_{a_3} \eta_{\alpha_3 \alpha_4} \big)
\nonumber \\ & \quad 
+ \big( \delta_{a_1a_2}\delta_{\alpha_1 \alpha_2}^> \xi_{a_1} +\eta_{a_1a_2}
 \delta_{\alpha_1 \alpha_2}^< \big) 
\big(\delta_{a_3a_4}\delta_{\alpha_3 \alpha_4}^> \xi_{a_3} +\eta_{a_3a_4}
 \delta_{\alpha_3 \alpha_4}^< \big) \Big] \mu(k_{12}) 
\nonumber
  \\ & \quad
+ \delta_{a_1a_2} \delta_{a_3a_4}
  \delta_{\alpha_1\alpha_2}  \delta_{\alpha_3\alpha_4} [\lambda(k_{12})
  +32\kappa U_\Lambda^{\prime\prime\prime} (\rho=2\kappa) 
  (\delta_{a_11}\delta_{\alpha_11} + \delta_{a_31}\delta_{\alpha_31})
\nonumber 
  \\ & \quad \quad \quad \quad
  +128 \kappa^2 U_\Lambda^{\prime\prime\prime\prime} (\rho=2\kappa)
  \delta_{a_11} \delta_{a_32} \delta_{\alpha_11}  \delta_{\alpha_31}]
  \nonumber \\ & \quad
+ [\delta_{\alpha_1 \alpha_2} - \Theta(\alpha_1\leq 2) \Theta(\alpha_2\leq 2) ]
[\delta_{\alpha_3 \alpha_4} - \Theta(\alpha_3\leq 2) \Theta(\alpha_4\leq 2) ]
\epsilon_{a_1 a_2} \epsilon_{a_3 a_4} [\Omega(k_{14})-\Omega(k_{13})] 
\nonumber \\
& \quad
  + (1\leftrightarrow 3) + (1\leftrightarrow 4)  \Big\} \, .
\end{align}
\end{widetext}

\section{Field expansion of local potential}
\label{app:fieldexpansion}
\begin{figure}[h]
\includegraphics[width=8.cm,clip,angle=0]{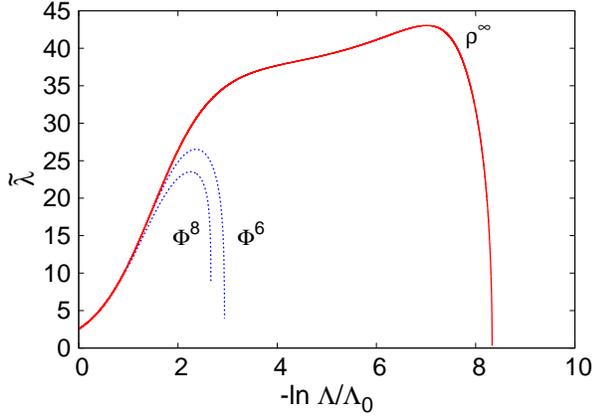}
\caption{(Color online) 
Flow of $\tilde{\lambda}$ at $T=0.275 J$ within different 
approximations. Shown are the results of a field expansion
to order $\Phi^6$ and $\Phi^8$ (dashed, blue lines) compared
to the full dependence on $\rho$ to first order in $\tau$ (red, solid line).}
\label{fig:compareLa}
\end{figure}

\begin{figure}[h]
\includegraphics[width=8.cm,clip,angle=0]{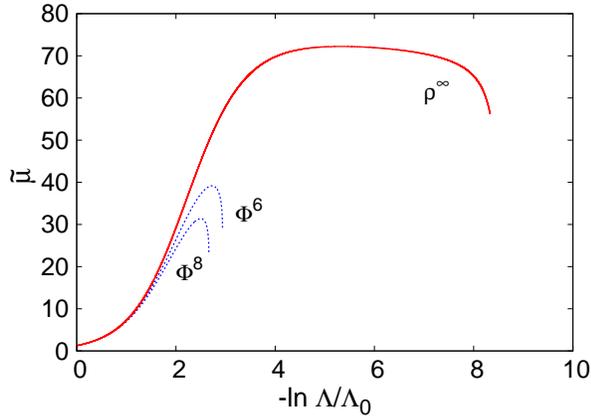}
\caption{(Color online) Flow of $\tilde{\mu}$ at $T=0.275 J$ within
 a field expansion
to order $\Phi^6$ and $\Phi^8$ (dashed, blue lines) compared
to the full dependence on $\rho$ to first order in $\tau$ (red, solid line).}
\label{fig:compareMu}
\end{figure}

\begin{figure}[h]
\includegraphics[width=8.cm,clip,angle=0]{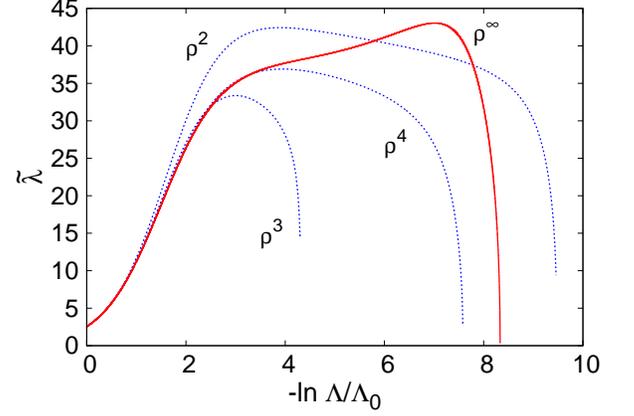}
\caption{(Color online) Flow of $\tilde{\lambda}$ at $T=0.275 J$ within an expansion
in $\rho$
to order $\rho^2$, $\rho^3$, and $\rho^4$ (dashed, blue lines) compared
to the full dependence on $\rho$ (red, solid line). All truncations keep only the first order 
in $\tau$.}
\label{fig:convergeLa}
\end{figure}

Here we discuss the flow equations if we approximate
the local potential $U_\Lambda(\rho,\tau)$ up to eighth order
in the fields. To that order, we have, up to a field independent
constant,
\begin{align}
\label{eq:expU}
  U_\Lambda(\tau,\rho/2-\kappa) & =\int_x \Big[ \frac{\lambda_\Lambda^0}{4}
  (\rho/2-\kappa)^2+\frac{\mu_\Lambda^0}{4} \tau+\frac{c_\rho^{(3)}}{12}
  (\rho/2-\kappa)^3 \nonumber \\
 &  +\frac{c_{\rho\tau}}{8}
  (\rho/2-\kappa) \tau+ \frac{c_\rho^{(4)}}{24} (\rho/2-\kappa)^4 \nonumber \\
& +\frac{c_\tau^{(2)}}{32} \tau^2 + \frac{c_{\rho\tau}^{(2)}}{16} (\rho/2-\kappa)^
2 \tau
\Big]
\end{align}
Higher order terms can be readily included, but the resulting
flow equations become rather long if the full $k$ dependence
of the coupling functions $\mu_\Lambda(k)$, $\lambda_\Lambda(k)$ and
$\Omega_\Lambda(k)$ is kept. To compare the different approximations,
we introduce the rescaled and (in $d=2$) dimensionless 
local coupling parameters 
\begin{subequations}
\begin{align}
\tilde{\mu}& = \mu_\Lambda^{(0)} \Lambda^{-2} Z_\Lambda^{-2} \, \, , 
\\
\tilde{\lambda} & = \lambda_\Lambda^{(0)} \Lambda^{-2} Z_\Lambda^{-2} \, \, .
\end{align}
\end{subequations}
In Fig.~\ref{fig:compareLa} we show 
the flow of $\tilde{\lambda}$ from the field expansion to both order
$\Phi^6$ (setting $c_\rho^{(4)}$, $c_\tau^{(2)}$ and $ c_{\rho\tau}^{(2)}$
equal to zero in Eq.~(\ref{eq:expU}) ) and to order $\Phi^8$. In
both cases $\tilde{\lambda}$ is driven rapidly to zero by a divergence
of a higher order vertex, much faster than in the approximation
given by Eq.~(\ref{eq:Local}) where all powers of $\rho$ are kept.
The same behavior is also observed in the flow of  $\tilde{\mu}$,
see Fig.~\ref{fig:compareMu}. In the $\Phi^8$ approximation
the suppression is even faster than in the $\Phi^6$ truncation
and what limits the flow is not the vanishing of the order
parameter $\rho^0$ but the divergence of the higher order
vertices. This clearly shows that a fixed order field expansion
is not useful in this case. In Fig.~\ref{fig:convergeLa}
we show that in contrast the expansion on just the
invariant $\rho$ shows better convergence properties.
The best alternative would be to
directly explore the flow of $U_\Lambda(\rho,\tau)$ without
any restrictions, which would however be numerically very
costly. As we discuss now, the higher order derivative
terms are also important, one would thus have to analyze the full flow 
of $U_\Lambda(\rho,\tau)$ in conjunction with the full
momentum dependence of the coupling functions $\mu_\Lambda(k)$,
$\lambda_\Lambda(k)$ and $\Omega_\Lambda(k)$, or at least with including
also higher order derivative terms,
which is numerically 
extremely challenging.

\begin{figure}[h]
\includegraphics[width=8.cm,clip,angle=0]{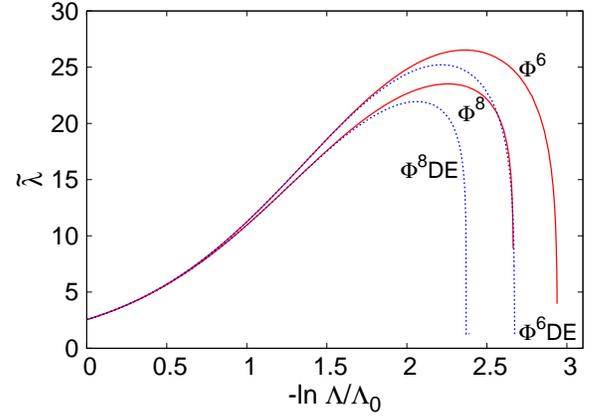}
\caption{(Color online)
Flow of $\tilde{\lambda}$ at $T=0.275 J$ with (red, full lines)
and without (dashed, blue lines) higher order derivative terms (see text,
DE stands for first order derivative expansion). Shown are results
including all local terms up to order $\Phi^6$ and to order $\Phi^8$.
}
\label{fig:compareDE}
\end{figure}

To gauge the importance of the $k$-dependent vertices,
we finally compare the flow within an approximation
where only the leading order derivative terms present in 
Eq.~(\ref{eq:EffActDE}) are kept with the approximation
where the full momentum dependence of $\lambda_\Lambda(k)$, $\mu_\Lambda(k)$
and $\Omega_\Lambda(k)$ is included. In both approximations
all local terms up to order $\Phi^6$ or $\Phi^8$ are included.
As shown in Fig.~\ref{fig:compareDE}, there are clear differences
and the flow with the full momentum dependence is more stable.
It thus seems that the higher order derivative terms are
non-negligible.

\end{document}